\documentclass[12pt]{article}
\usepackage[latin9]{inputenc}
\usepackage{color}
\usepackage{array}
\usepackage{float}
\usepackage{units}
\usepackage{multirow}
\usepackage{amsmath}
\usepackage{amssymb}
\usepackage{cancel}
\usepackage{graphicx}
\usepackage{rotfloat}

\makeatletter

\providecommand{\tabularnewline}{\\}

\usepackage{ulem}
\usepackage{cite}
\usepackage{rotating}
\usepackage{amsfonts}
\usepackage{multirow}

\newcommand{\beq}{\begin{equation}}
\newcommand{\eeq}{\end{equation}}

\makeatother

\begin{document}
\begin{titlepage}

\vskip 2cm 
\begin{center}
\textbf{\Large{}{}{}Non-commutativity and non-inertial effects on
a scalar field in a cosmic string space-time}{\Large{}}\footnote{\texttt{Email: rodrigo.cuzinatto@unifal-mg.edu.br, mdemonti@ualberta.ca,
pompeia@ita.br}}{\Large{}}\\
{\Large{} {} }\textbf{\Large{}{}{}Part 2: Spin-zero Duffin-Kemmer-Petiau-like
oscillator}{\Large\par}
\par\end{center}

\begin{center}
 \par\end{center}

\begin{center}
\textbf{\Large{}}\textbf{ \vskip 10pt Rodrigo Rocha Cuzinatto$^{a}$, Marc de Montigny$^{b}$,
Pedro Jos\'e Pompeia$^{c}$ } \vskip 5pt \textsl{$^{a}$Instituto de
Ci\^encia e Tecnologia, Universidade Federal de Alfenas}\\
\textsl{ Rodovia Jos\'e Aur\'elio Vilela, 11999, Cidade Universit\'aria}\\
\textsl{ CEP 37715-400 Po\c cos de Caldas, Minas Gerais, Brazil} \vskip
2pt \textsl{$^{b}$Facult\'e Saint-Jean, University of Alberta }\\
\textsl{ 8406 91 Street NW}\\
\textsl{ Edmonton, Alberta, Canada T6C 4G9, Canada} \vskip 2pt \textsl{$^{c}$Departamento
de F\'{\i}sica, Instituto Tecnol\'ogico de Aeron\'autica}\\
\textsl{ Pra\c ca Mal. Eduardo Gomes 50}\\
\textsl{ CEP 12228-900 S\~ao Jos\'e dos Campos, S\~ao Paulo, Brazil} 
\par\end{center}

\begin{abstract}
We study the non-inertial effects of a rotating frame on a spin-zero,
Duffin-Kemmer-Petiau (DKP)-like oscillator in a cosmic string space-time
with non-commutative geometry in the momentum space. The spin-zero
DKP-like oscillator is obtained from the Klein-Gordon Lagrangian with
a non-standard prescription for the oscillator coupling. We find that
the solutions of the time-independent radial equation with the non-zero
non-commutativity parameter parallel to the string are related to
the confluent hypergeometric function. We find the quantized energy
eigenvalues of the non-commutative oscillator. 
\end{abstract}
\bigskip{}


\bigskip{}

\end{titlepage}

\newpage{}

\section{Introduction\label{introduction}}

A common approach to the theory of spin-zero fields involves second-order
differential equations of motion, such as the Klein-Gordon (KG) equation
(e.g. see Ref. \cite{Weinberg}). The Duffin-Kemmer-Petiau (DKP) equation
\cite{Duffin,Kemmer,Petiau,Greiner} was an attempt to obtain first-order
field equations for fields with integer spin analogous to the Dirac
equation for spin-half fields.  Both descriptions are not always equivalent
and, for this reason, it is key to characterize the phenomenology 
associated both to KG field and DKP field so that experimental results
can ultimately decide in favour of one description over the other \cite{Friedman1986,YangHassanabadi2021}.

Even though the equivalence between the DKP and KG equations for interacting
spin-zero fields is not absolute \cite{Lunardi}, at least, it is
well known that both are equivalent for free fields. For interacting
fields, a coupling prescription is required and it can be introduced
either by symmetry or by phenomenological arguments. We should keep
in mind that different prescriptions lead to different results, and
each case should be analyzed separately \cite{Utiyama}. The DKP equation
lends itself to a richness of couplings that are not all equivalent
to KG theories \cite{Guertin,Vijayalakshmi}. The fundamental point
in this regard is to address whether it is possible to obtain a second-order
differential equation for a spin-zero field that leads to an equivalent
DKP-oscillator equation. In this paper, we will consider a system
for which it is indeed possible to obtain such equation. However,
there will be a price to pay: the oscillator coupling has to be different
from the ordinary prescription. Indeed, as we did in Ref. \cite{CQG1},
the standard oscillator prescription consists of taking the non-minimal
coupling 
\begin{equation}
\begin{split} & \mathbf{p}\phi\rightarrow\ \left(\mathbf{p}-im\omega_{0}\mathbf{r}\right)\phi,\\
 & \mathbf{p}\phi^{*}\rightarrow\ \left(\mathbf{p}+im\omega_{0}\mathbf{r}\right)\phi^{*},
\end{split}
\label{nonminimal_KG}
\end{equation}
where $m$ is the mass of the complex scalar field $\phi$ localized
by coordinate $\mathbf{r}$ and momentum $\boldsymbol{{\rm p}}$,
and $\omega_{0}$ is the oscillator's frequency. This leads to the
so-called `KG-oscillator'.

Hereafter, we shall take a different path and consider a similar prescription
but without modifying the sign of the operator $\mathbf{r}$, as we
did in Eq. \eqref{nonminimal_KG} for the field and its conjugate.
We will see that that second path leads to a second-order equation
which is equivalent to the DKP-oscillator. We intend to explore the
role of this alternative prescription upon the behavior of a spin-zero
field oscillator. Physically, different couplings may lead to distinct
effects through the interaction. Our goal is to compare the results
herein with those obtained in Ref. \cite{CQG1}, hence referred to
as Part 1. 

In Part 1, besides considering the oscillator prescription, we also
took into account the effects of non-commutativity in momentum space
via 
\begin{equation}
\begin{split} & \mathbf{p}\phi\rightarrow\ \left(\mathbf{p}-\frac{\boldsymbol{\Omega}\times\mathbf{r}}{2}\right)\phi,\\
 & \mathbf{p}\phi^{*}\rightarrow\ \left(\mathbf{p}+\frac{\boldsymbol{\Omega}\times\mathbf{r}}{2}\right)\phi^{*},
\end{split}
\label{nonComm_KG}
\end{equation}
where $\boldsymbol{\Omega}$ is the non-commutative vector parameter.
The sign choices in Eqs.(\ref{nonminimal_KG}) and (\ref{nonComm_KG})
are clearly similar. Analogously, in this work we will demand consistency
between the choice of sign in the DKP-like oscillator prescription
and the choice of sign in the recipe for introducing non-commutativity.

We will analyze the spin-zero DKP-like oscillator in a cosmic string
spacetime, which, in a rotating frame, is described by the line element
\begin{equation}
ds^{2}=-\left(1-\omega^{2}\eta^{2}\rho^{2}\right)dt^{2}+2\omega\eta^{2}\rho^{2}d\varphi dt+d\rho^{2}+\eta^{2}\rho^{2}d\varphi^{2}+dz^{2}.\label{CosmicString}
\end{equation}
We use cylindrical coordinates $(\rho,\varphi,z)$, where $\rho\in[0,\infty),\,\varphi\in[0,2\pi],\,z\in(-\infty,+\infty)$,
and natural units where $c=\hbar=1$. The angular frequency of the
rotating frame is given by $\omega$; the string parameter is $\eta=1-4\Lambda\in(0,1]$,
where $\Lambda$ is the topological defect linear mass density (see Refs. \cite{Hassanabadi2016,Hosseinpour2019,Chen2020,Yang2021}
for other applications). Notice that this line element is singular
at the hard-wall 
\begin{equation}
\rho_{0}\equiv\frac{1}{\omega\eta}.\label{hardwall}
\end{equation}
As a consequence, our particle should be confined within $0\leqslant\rho<\rho_{0}$,
which means that the wave function has to vanish at the hard-wall.
Hereafter, we shall consider a similar situation as
in Part 1, where we examined a KG-oscillator in the presence of non-commutativity
in a rotating frame in a cosmic string spacetime \cite{CQG1}. Therein,
we observed that non-commutativity was responsible for changing the
energy eigenstates of the particle both when the hard-wall is at a
finite position and when the hard-wall's position approaches infinity.
There, we saw that we can interpret the non-commutativity parameter
as a constant magnetic field pointing in the direction of the string.
Hereafter, a similar interpretation does not hold due to a difference
of sign in the coupling, cf. Section \ref{SecCosmicString}. Even
without that connection between non-commutativity and a magnetic field,
in this paper, we are able to describe a spin-zero DKP-like oscillator.
More details about the DKP oscillator can be found in Refs. \cite{Castro,HosseinpourHassanabadiAndrade2018,HosseinpourHassanabadi2018}
and references therein. Moreover, we can shed light on how the coupling
between the field and non-commutativity could be carried out. Recall
that in the case of the electromagnetic interaction, the coupling
between the scalar field and the electromagnetic potential is determined
by the minimal coupling prescription, which is a consequence of the
underlying $U\left(1\right)$ gauge symmetry. With non-commutativity,
this coupling is not connected with any gauge symmetry and could be
introduced in at least two different ways: one is presented in Eq.
(\ref{nonComm_KG}) and was explored in Part 1, in which the focus
was on the mapping between non-commutativity and a magnetic field;
the other is to apply $\mathbf{p}\rightarrow\ \mathbf{p}-\frac{\boldsymbol{\Omega}\times\mathbf{r}}{2}$
to both $\phi$ and $\phi^{*}$. The latter possibility is the one
explored here and we focus on the effects of the different choice
of coupling prescription for the non-commutativity upon the physical
system.

In Section \ref{SecCosmicString}, we define the KG Lagrangian and
establish the field equation for the spin-zero DKP-like oscillator
in a non-inertial frame and cosmic string space-time with non-commutative
geometry. We thus obtain an equation in terms of time and cylindrical
coordinates, which we solve in Section \ref{subsec:Solution} with
the only non-zero non-commutativity parameter parallel to the cosmic
string. The solution along that axis is in terms of the Hermite polynomials,
whereas the radial solutions are expressed in terms of the confluent
hypergeometric functions. We discuss the corresponding energy eigenvalues
in Section \ref{Sec:Energy}: with the hard-wall at infinity in Section
\ref{SubSec:HDinf} and at a finite distance from the cosmic string
in Section \ref{SubSec:HDfinite}. We complete the paper with concluding
remarks in Section \ref{Sec:Conclusion}.


\section{Spin-zero DKP-like oscillator in a non-inertial frame and cosmic
string space-time with non-commutative geometry\label{SecCosmicString}}

The equation for the DKP-like oscillator is obtained from the KG equation
with the non-minimal coupling \cite{GERG2019,Moshinsky,Bakke,Nedjadi}:
\begin{equation}
\mathbf{p}\rightarrow\ \mathbf{p}-im\omega_{0}\mathbf{r},\label{nonminimal}
\end{equation}
where we use cylindrical coordinates and select 
\begin{equation}
\mathbf{r}=\rho\hat{e}_{\rho}+z\hat{e}_{z}=\left(\rho,0,z\right).
\end{equation}
Since $p_{j}=-{\rm i}\partial_{j}$ ($j=1,2,3$), Eq. (\ref{nonminimal})
is $-{\rm i}\left(\nabla\pm m\omega_{0}\mathbf{r}\right)$.

If we begin with the KG Lagrangian 
\begin{equation}
L=\sqrt{-g}\left(-g^{ab}\partial_{a}\phi^{*}\partial_{b}\phi-m^{2}\phi^{*}\phi\right),
\end{equation}
and perform the non-minimal substitution of Eq. \eqref{nonminimal},
we obtain the Lagrangian of the KG oscillator: 
\begin{eqnarray}
L_{\mathrm{osc}} & = & -\sqrt{-g}\left[g^{00}\partial_{0}\phi^{*}\partial_{0}\phi+g^{02}\partial_{0}\phi^{*}\partial_{2}\phi+g^{02}\partial_{2}\phi^{*}\partial_{0}\phi\right.\nonumber \\
 &  & +g^{11}\left(\partial_{1}\phi^{*}+m\omega_{0}\rho\phi^{*}\right)\left(\partial_{1}\phi+m\omega_{0}\rho\phi\right)+g^{22}\partial_{2}\phi^{*}\partial_{2}\phi\label{Lag-osc}\\
 &  & \left.+g^{33}\left(\partial_{3}\phi^{*}+m\omega_{0}z\phi^{*}\right)\left(\partial_{3}\phi+m\omega_{0}z\phi\right)+m^{2}\phi^{*}\phi\right],\nonumber 
\end{eqnarray}
which, with 0, 1, 2, 3 standing for $t$, $\rho$, $\varphi$ and
$z$, respectively, leads to the equation of motion 
\begin{equation}
-m^{2}\phi-m^{2}\omega_{0}^{2}\rho^{2}\phi-m^{2}\omega_{0}^{2}z^{2}\phi-\partial_{0}^{2}\phi+\frac{1}{\rho}\partial_{1}\phi+\partial_{1}^{2}\phi+\left(\frac{1}{\eta^{2}\rho^{2}}-\omega^{2}\right)\partial_{2}^{2}\phi+2\omega\partial_{2}\partial_{0}\phi+\partial_{3}^{2}\phi+3m\omega_{0}\phi=0.\label{RodEq4}
\end{equation}

Next we introduce the non-commutative momentum space. Consider a non-commutative
phase space described by the operators $\hat{r}_{i}$ and $\hat{p}_{i}$,
\begin{align}
\hat{r}_{i} & =r_{i}-\frac{\Theta_{ij}}{2\hbar}p_{j}=r_{i}+\frac{\left(\mathbf{\Theta}\times{\bf p}\right)_{i}}{2\hbar},\label{transformation-nc-r}\\
\hat{p}_{i} & =p_{i}+\frac{\Omega_{ij}}{2\hbar}r_{j}=p_{i}-\frac{\left(\mathbf{\Omega}\times{\bf r}\right)_{i}}{2\hbar},\label{transformation-nc-p}
\end{align}
which satisfy the following commutation relations: 
\begin{equation}
\left[\hat{r}_{i},\hat{r}_{j}\right]={\rm i}\Theta_{ij},\quad\left[\hat{p}_{i},\hat{p}_{j}\right]={\rm i}\Omega_{ij},\quad\left[\hat{r}_{i},\hat{p}_{j}\right]={\rm i}\hbar\Delta_{ij},
\end{equation}
where $\Theta_{ij}=\epsilon_{ijk}\Theta_{k}$, $\Omega_{ij}=\epsilon_{ijk}\Omega_{k}$,
with $\Theta_{i}$ and $\Omega_{i}$ ($i=1,2,3$) real parameters.
The matrix $\Delta_{ij}$ is given by 
\begin{equation}
\Delta_{ij}=\left(1+\frac{\mathbf{\Theta}\cdot\mathbf{\Omega}}{4\hbar^{2}}\right)\delta_{ij}-\frac{\Omega_{i}\Theta_{j}}{4\hbar^{2}}.
\end{equation}

From the onset, we will restrict our study to 
\begin{equation}
\Theta_{j}=0,\quad j=1,2,3,
\end{equation}
so that we allow for a non-commutative momentum space only, whereas
the configuration space for the cosmic string remains commutative
for reasons discussed in Paper 1. Therefore, the non-commutative phase-space
components are 
\begin{align}
 & \hat{r}_{i}=r_{i},\nonumber \\
 & \hat{p}_{i}=p_{i}+\frac{\Omega_{ij}}{2}r_{j}=p_{i}-\frac{\left(\mathbf{\Omega}\times{\bf r}\right)_{i}}{2},\label{pNC}
\end{align}
so that the commutation relations read 
\begin{eqnarray}
\left[\hat{r}_{i},\hat{r}_{j}\right]=0,\quad\left[\hat{p}_{i},\hat{p}_{j}\right]={\rm i}\Omega_{ij},\quad\left[\hat{r}_{i},\hat{p}_{j}\right]={\rm i}\hbar\delta_{ij}.
\end{eqnarray}
In cylindrical coordinates, the second term of Eq. (\ref{pNC}) is
\begin{equation}
\frac{\Omega_{ij}}{2}r_{j}=\frac{1}{2}\left(\Omega_{13}z,\Omega_{21}\rho+\Omega_{23}z,\Omega_{31}\rho\right)=\frac{1}{2}\left(-\Omega_{2}z,-\Omega_{3}\rho+\Omega_{1}z,\Omega_{2}\rho\right).
\end{equation}

Then, Eq. (\ref{Lag-osc}) is generalized to the non-commutative oscillator
Lagrangian: 
\begin{eqnarray}
L_{\mathrm{NC-osc}} & = & -\sqrt{-g}\left[g^{00}\partial_{0}\phi^{*}\partial_{0}\phi+g^{02}\partial_{0}\phi^{*}\left(\partial_{2}\phi+i\frac{1}{2}\Omega_{3}\rho\phi-\frac{1}{2}i\Omega_{1}z\phi\right)\right.\nonumber \\
 &  & +g^{02}\left(\partial_{2}\phi^{*}+\frac{1}{2}i\Omega_{3}\rho\phi^{*}-\frac{1}{2}i\Omega_{1}z\phi^{*}\right)\partial_{0}\phi\nonumber \\
 &  & +g^{11}\left(\partial_{1}\phi^{*}+m\omega_{0}\rho\phi^{*}+\frac{1}{2}i\Omega_{2}z\phi^{*}\right)\left(\partial_{1}\phi+m\omega_{0}\rho\phi+\frac{1}{2}i\Omega_{2}z\phi\right)\nonumber \\
 &  & +g^{22}\left(\partial_{2}\phi^{*}+\frac{1}{2}i\Omega_{3}\rho\phi^{*}-\frac{1}{2}i\Omega_{1}z\phi^{*}\right)\left(\partial_{2}\phi+i\frac{1}{2}\Omega_{3}\rho\phi-\frac{1}{2}i\Omega_{1}z\phi\right)\nonumber \\
 &  & \left.+g^{33}\left(\partial_{3}\phi^{*}+m\omega_{0}z\phi^{*}-\frac{1}{2}i\Omega_{2}\rho\phi^{*}\right)\left(\partial_{3}\phi+m\omega_{0}z\phi-\frac{1}{2}i\Omega_{2}\rho\phi\right)+m^{2}\phi^{*}\phi\right],
\end{eqnarray}
which modifies Eq. (\ref{RodEq4}) as follows: 
\begin{equation}
\begin{array}{ccc}
-m^{2}\phi+3m\omega_{0}\phi-\left(m\omega_{0}\rho+i\frac{1}{2}\Omega_{2}z\right)^{2}\phi-\left(m\omega_{0}z-i\frac{1}{2}\Omega_{2}\rho\right)^{2}\phi\\
-\left(\frac{1}{\eta^{2}\rho^{2}}-\omega^{2}\right)\left(i\frac{1}{2}\Omega_{3}\rho-i\frac{1}{2}\Omega_{1}z\right)^{2}\phi+i\frac{1}{2}\Omega_{2}\frac{z}{\rho}\phi\\
+\frac{1}{\rho}\partial_{1}\phi+2\omega\partial_{0}\partial_{2}\phi-\partial_{0}^{2}\phi+\partial_{1}^{2}\phi+\left(\frac{1}{\eta^{2}\rho^{2}}-\omega^{2}\right)\partial_{2}^{2}\phi+\partial_{3}^{2}\phi=0.
\end{array}\label{EMNC}
\end{equation}
In the next section, we will further simplify our problem with $\Omega_{3}$
being the only non-zero non-commutativity parameter.


\subsection{Solution to the (momentum-space) non-commutative field equation \label{subsec:Solution}}

Now we look for the corresponding time-independent solution $\psi(\rho,\varphi,z)$
of Eq.~(\ref{EMNC}), defined by 
\begin{equation}
\phi\left(\rho,\varphi,z,t\right)=e^{-i{\cal E}t}\psi\left(\rho,\varphi,z\right),
\end{equation}
where ${\cal E}$ is the energy, so that Eq. (\ref{EMNC}) is reduced
to 
\begin{eqnarray}
-m^{2}\psi+3m\omega_{0}\psi+\left(\frac{1}{4}\Omega_{2}^{2}-m^{2}\omega_{0}^{2}\right)\left(\rho^{2}+z^{2}\right)\psi+\left(\frac{1}{\eta^{2}\rho^{2}}-\omega^{2}\right)\left(\frac{1}{2}\Omega_{3}\rho-\frac{1}{2}\Omega_{1}z\right)^{2}\psi\nonumber \\
+\frac{1}{2}i\Omega_{2}\frac{z}{\rho}\psi+\frac{1}{\rho}\partial_{\rho}\psi-2i\omega\mathcal{E}\partial_{\varphi}\psi+{\cal E}^{2}\psi+\partial_{\rho}^{2}\psi+\left(\frac{1}{\eta^{2}\rho^{2}}-\omega^{2}\right)\partial_{\varphi}^{2}\psi+\partial_{z}^{2}\psi=0.\label{EMNC-t}
\end{eqnarray}
We note that the non-commutativity parameters {${\bf {\Omega}}$}
only couple to the frame angular velocity $\omega$ of Eq. (\ref{CosmicString}),
so that there is no coupling with the oscillator frequency $\omega_{0}$
of Eq. (\ref{nonminimal}). Moreover, we observe that the hard-wall
condition is manifest through the factors $\frac{1}{\eta^{2}\rho^{2}}-\omega^{2}$
in Eq. \eqref{EMNC-t}, as it was in Eqs. \eqref{RodEq4} and \eqref{EMNC}.

At this point, if, in order to find a solution, we use the method
of separation of variables, we can observe from Eq. (\ref{EMNC-t})
that this method is not easily applicable, since there are terms that
involves the product of coordinates $\rho$ and $z$: $\frac{1}{2}i\Omega_{2}\frac{z}{\rho}\psi$
and $\left(\frac{1}{\eta^{2}\rho^{2}}-\omega^{2}\right)\left(\frac{1}{2}\Omega_{3}\rho-\frac{1}{2}\Omega_{1}z\right)^{2}\psi$.
We avoid this problem by selecting the particular case, 
\begin{equation}
\Omega_{1}=\Omega_{2}=0,\label{Om1Om2}
\end{equation}
so that we take $\Omega_{3}\neq0$ only. This particular choice is
not only convenient from the computational stand point, but also physically
meaningful as shown in the discussion above on the mapping of momentum-space
non-commutativity with $\Omega_{3}\neq0$ to a constant magnetic field
pointing in the $z$-direction \cite{Delduc}. We implement the separation
of variables: 
\begin{equation}
\psi\left(\rho,\varphi,z\right)=R\left(\rho\right)\Phi\left(\varphi\right)Z\left(z\right),\label{separation}
\end{equation}
which we substitute in Eq. (\ref{EMNC-t}) and obtain 
\begin{eqnarray}
{\cal E}^{2}-m^{2}+3m\omega_{0}+\left[\left(\frac{1}{\eta^{2}\rho^{2}}-\omega^{2}\right)\frac{1}{4}\Omega_{3}^{2}-m^{2}\omega_{0}^{2}\right]\rho^{2}+\frac{1}{\rho}\frac{1}{R\left(\rho\right)}\partial_{\rho}R\left(\rho\right)+\frac{1}{R\left(\rho\right)}\partial_{\rho}^{2}R\left(\rho\right)\nonumber \\
-m^{2}\omega_{0}^{2}z^{2}+\frac{1}{Z\left(z\right)}\partial_{z}^{2}Z\left(z\right)-2i\omega\mathcal{E}\frac{1}{\Phi\left(\varphi\right)}\partial_{\varphi}\Phi\left(\varphi\right)+\left(\frac{1}{\eta^{2}\rho^{2}}-\omega^{2}\right)\frac{1}{\Phi\left(\varphi\right)}\partial_{\varphi}^{2}\Phi\left(\varphi\right)=0.\label{EM1}
\end{eqnarray}

In order to further decouple this equation, we utilize the azimuthal
coordinate ansatz: 
\begin{equation}
\Phi=e^{iL\varphi},\label{eq:AnsatzPhi}
\end{equation}
which is subject to the familiar boundary condition $\Phi\left(2\pi\right)=\Phi\left(0\right)$,
so that $e^{iL\left(2\pi\right)}=1$, which leads to the quantization
of the number $L$ in Eq. (\ref{eq:AnsatzPhi}): 
\begin{equation}
L=0,\pm1,\pm2,\pm3,\cdots\label{eq:QuantizationL}
\end{equation}

Let us set the $z$-dependent part equal to a constant, 
\begin{equation}
\frac{1}{Z}\left[\partial_{z}^{2}Z-\left(m\omega_{0}\right)^{2}z^{2}Z\right]=-2mk,\label{equ23}
\end{equation}
where $m$ is the mass of the Klein-Gordon field, and $k$ a new constant
with units of energy. We replace the variable $z$ with the dimensionless
variable $\zeta$, 
\begin{equation}
\zeta\equiv\sqrt{m\omega_{0}}z.\label{eq:zeta}
\end{equation}
Then Eq. (\ref{equ23}) is simplified to 
\begin{equation}
\frac{d^{2}Z}{d\zeta^{2}}=\left[\zeta^{2}-\frac{2k}{\omega_{0}}\right]Z=\left[\zeta^{2}-K\right]Z,\label{equ25}
\end{equation}
where 
\begin{equation}
K\equiv\frac{2k}{\omega_{0}}.\label{eq:K}
\end{equation}
This occurrence of an energy unit of the quantum harmonic oscillator
manifests the oscillatory nature of the model. This is also shown
by the fact that Eq. (\ref{equ25}) is exactly the harmonic oscillator
differential equation (see, e.g. Ref. \cite{Griffiths}). Thus the
solutions for $Z(\zeta)$ are given in terms of Hermite polynomials:
\begin{equation}
Z\left(\zeta\right)=\left(\frac{m\omega_{0}}{\pi}\right)^{1/4}\frac{1}{\sqrt{2^{m_{z}}m_{z}!}}H_{m_{z}}\left(\zeta\right)e^{-\zeta^{2}/2},\label{eq:Z(zeta)}
\end{equation}
where $m_{z}=0,1,2,\cdots$ Note also that the recursion formula of
the Frobenius expansion requires $K=2m_{z}+1$, which, from Eq. (\ref{eq:K}),
leads to the quantization of $k$ 
\begin{equation}
k=\omega_{0}\left(m_{z}+\frac{1}{2}\right)\,,\qquad m_{z}=0,1,2,...\label{eq:Quantization_k}
\end{equation}
Although the normalization factor in Eq. (\ref{eq:Z(zeta)}) is not
particularly relevant at this point, we choose it such that the coefficient
of the highest power in $\zeta$ is $2^{m_{z}}$, as in Ref. \cite{Schiff}.

Finally, let us turn to the most intricate part of Eq. (\ref{EM1}):
the radial part, which reads 
\begin{equation}
\frac{d^{2}R}{d\rho^{2}}+\frac{1}{\rho}\frac{dR}{d\rho}-\left[\left(m\omega_{0}\right)^{2}+\left(\frac{\Omega_{3}}{2}\omega\right)^{2}\right]\rho^{2}R-\left(\frac{L}{\eta}\right)^{2}\frac{1}{\rho^{2}}R+\left[\left(\mathcal{E}+\omega L\right)^{2}-m^{2}+2m\omega_{0}\left(1-m_{z}\right)+\left(\frac{\Omega_{3}}{2\eta}\right)^{2}\right]R=0.\label{eq:DiffEqR(rho)}
\end{equation}
Instead of $\rho$, we use as an independent-variable a dimensionless
radial coordinate $\xi$: 
\begin{equation}
\xi\equiv S\rho,\label{eq:xi}
\end{equation}
where 
\begin{equation}
S=S\left(m,\omega_{0},\omega,\Omega_{3}\right)\equiv\left[\left(m\omega_{0}\right)^{2}+\left(\frac{\Omega_{3}}{2}\omega\right)^{2}\right]^{\frac{1}{4}}.\label{eq:S}
\end{equation}
Then Eq. (\ref{eq:DiffEqR(rho)}) is reduced into 
\begin{equation}
\frac{d^{2}R}{d\xi^{2}}+\frac{1}{\xi}\frac{dR}{d\xi}-\xi^{2}R-\left(\frac{L}{\eta}\right)^{2}\frac{1}{\xi^{2}}R+\left(\frac{T}{S}\right)^{2}R=0,\label{eq:DiffEqR(xi)}
\end{equation}
where 
\begin{equation}
T=T\left(\mathcal{E},m,\omega_{0},\omega,\Omega_{3}\right)\equiv\left[\left(\mathcal{E}+\omega L\right)^{2}-m^{2}+2m\omega_{0}\left(1-m_{z}\right)+\left(\frac{\Omega_{3}}{2\eta}\right)^{2}\right]^{\frac{1}{2}}\label{eq:T}
\end{equation}
is a constant, with respect to the radial coordinate $\xi$ (and $\rho$).

In Section \ref{Sec:Energy}, we will properly consider the hard-wall
condition restricting $\rho$ (and $\xi$) to a finite value. For
now, we shall solve Eq.~(\ref{eq:DiffEqR(xi)}) with an ansatz suggested
by the asymptotic behaviour of the solution at infinity and near the
origin. Firstly, in the limit at infinity, $\xi\rightarrow\infty$,
the term scaling with $\xi^{2}$ dominates completely in Eq.~(\ref{eq:DiffEqR(xi)}),
so that we choose 
\begin{equation}
R\left(\xi\right)\approx Ae^{-\xi^{2}/2}\qquad\left(\xi\rightarrow\infty\right).\label{eq:R(xi)_large_xi}
\end{equation}
Secondly, the asymptotic behaviour near the origin, $\xi\rightarrow0$,
shows that the term proportional to $1/\xi^{2}$ in Eq. (\ref{eq:DiffEqR(xi)})
dominates so that the asymptotic behaviour is 
\begin{equation}
R\left(\xi\right)\approx C\xi^{q}\qquad\left(\xi\rightarrow0\right),\label{eq:R(xi)_small_xi}
\end{equation}
where 
\begin{equation}
q=\frac{|L|}{\eta}.\label{eq:q_plus}
\end{equation}
Therefore, we construct the ansatz used to solve Eq. (\ref{eq:DiffEqR(xi)})
by gluing the asymptotic behaviours shown in Eqs. (\ref{eq:R(xi)_large_xi})
and (\ref{eq:R(xi)_small_xi}) for $\xi\rightarrow0$, with a new
function $h\left(\xi\right)$: 
\begin{equation}
R\left(\xi\right)=\xi^{q}h\left(\xi\right)e^{-\xi^{2}/2}.\label{eq:R(xi)_asymp}
\end{equation}
We substitute Eq. (\ref{eq:R(xi)_asymp}) into Eq. (\ref{eq:DiffEqR(xi)})
and, after lengthy calculations, we obtain a differential equation
for $h(\xi)$, 
\begin{equation}
\xi^{2}\frac{d^{2}h}{d\xi^{2}}+2\left(q+\frac{1}{2}-\xi^{2}\right)\xi\frac{dh}{d\xi}+\left\{ \left[\upsilon_{0}^{2}-2\left(q+1\right)\right]\xi^{2}\right\} h=0,\label{eq:(xi)DiffEqh(xi)}
\end{equation}
where 
\begin{equation}
\upsilon_{0}^{2}\equiv\left(\frac{T}{S}\right)^{2}.\label{eq:xi0}
\end{equation}
($S$ and $T$ are defined in Eqs. \eqref{eq:S} and \eqref{eq:T},
respectively.)

Next we look for solutions to the second-order ordinary differential
equation (\ref{eq:(xi)DiffEqh(xi)}). We can see that the coefficients
in the Frobenius method correspond to a Kummer series, or confluent
hypergeometric function, 
\begin{equation}
_{1}F_{1}(a;b;\chi)=1+\frac{a}{b}\chi+\frac{a(a+1)}{b(b+1)}\frac{\chi^{2}}{2!}+\cdots\label{equ-Kummer}
\end{equation}
where 
\begin{equation}
a\equiv\left(1+q-\upsilon_{0}^{2}/2\right)/2,\label{equ-a}
\end{equation}
\begin{equation}
b\equiv\left(1+q\right),\label{equ-b}
\end{equation}
and 
\begin{equation}
\chi\equiv\xi^{2}.\label{equZ}
\end{equation}
In the next section, we will use this result to discuss the energy
eigenvalues for our system. This will be done in two ways: first,
we determine $\mathcal{E}$ in the limit of a hard-wall pushed to
infinity, which allows to obtain an analytical solution for $\mathcal{E}$
and second, the hard-wall at a finite distance, which is physically
more interesting, but requires numerical techniques.


\section{Energy eigenvalues\label{Sec:Energy}}


\subsection{Hard-wall at infinity\label{SubSec:HDinf}}

In this section, we consider the limit where the location $\rho_{0}\equiv1/\omega\eta$
of the hard-wall approaches infinity, which of course occurs when
at least one between rotating frame frequency $\omega$ or the cosmic
string parameter $\eta$ tends to zero.

The Frobenius series applied to Eq.~(\ref{eq:(xi)DiffEqh(xi)}) reduces
to a polynomial if there exists a maximum value of the summation index
$j$, say $j_{\text{max}}=2n$, where $n=0,1,2,3,...$ such that
$c_{j_{\text{max}}}$ is the coefficient of the highest power of $\xi^{2}$.
This amounts to demanding that the numerator of the associated recurrence
relation vanishes: 
\begin{equation}
\left[2\left(2n+1+q\right)-\upsilon_{0}^{2}\right]=0,\qquad j_{\text{max}}=2n.\label{eq:polynomial-cond}
\end{equation}
Then the quantization relation reads 
\begin{equation}
\upsilon_{0}^{2}=2\left(2n+1+\left|\frac{L}{\eta}\right|\right),\qquad n=0,1,2,\cdots\label{eq:Quantization(xi0,L)}
\end{equation}
where we used Eq.~(\ref{eq:q_plus}) for $q$. The energy $\mathcal{E}$
is contained within the quantity $\upsilon_{0}$, defined in Eq. (\ref{eq:xi0}),
which depends also on the non-commutative parameter $\Omega_{3}$
via $T$ (see Eq. (\ref{eq:T})) and $S$, in Eq. (\ref{eq:S}). From
Eq (\ref{eq:Quantization(xi0,L)}) and the definition of $\upsilon_{0}$,
\begin{equation}
\upsilon_{0}^{2}=\frac{\left(\mathcal{E}+\omega L\right)^{2}-m^{2}+2m\omega_{0}\left(1-m_{z}\right)+\left(\frac{\Omega_{3}}{2\eta}\right)^{2}}{\left[\left(m\omega_{0}\right)^{2}+\left(\frac{\Omega_{3}}{2}\omega\right)^{2}\right]^{\frac{1}{2}}},\label{eq:xi0(E)}
\end{equation}
if we solve for $\mathcal{E}$, then we obtain 
\begin{equation}
\mathcal{E}=-\omega L\pm\sqrt{m^{2}-2m\omega_{0}\left(1-m_{z}\right)-\left(\frac{\Omega_{3}}{2\eta}\right)^{2}+2\left(2n+1+\frac{\left|L\right|}{\eta}\right)\sqrt{\left(m\omega_{0}\right)^{2}+\left(\frac{\Omega_{3}}{2}\omega\right)^{2}}}\label{eq:E_pm}
\end{equation}
In the limit $\omega\rightarrow0$, $\omega_{0}\rightarrow0$, $\Omega_{3}\rightarrow0$,
and $\eta\rightarrow1$, Eq.~(\ref{eq:E_pm}) reduces to $\mathcal{E}=\pm\sqrt{m^{2}}$,
which suggests to keep the plus sign in (\ref{eq:E_pm}) in order
to describe particles. Alternatively, we could interpret the negative
sign in the second term as associated to antiparticles. Note, however,
that the term $-\omega L$, due to the non-inertial rotating frame,
prevents the antiparticles' energy levels to be simply equal to $-1$
times the particles' energy levels. Our subsequent analysis will focus
on particles solely, and can be readily adapted to antiparticles.

Therefore our energy eigenvalues for particles are 
\begin{equation}
\mathcal{E}_{nLm_{z}}=-\omega L+\sqrt{m^{2}-2m\omega_{0}\left(1-m_{z}\right)-\left(\frac{\Omega_{3}}{2\eta}\right)^{2}+2\left(2n+1+\frac{\left|L\right|}{\eta}\right)\sqrt{\left(m\omega_{0}\right)^{2}+\left(\frac{\Omega_{3}}{2}\omega\right)^{2}}},\label{eq:E}
\end{equation}
where $n=0,1,2,\dots$, $L=0,\pm1,\pm2,...$ and $m_{z}=0,1,2,\cdots$.
Note that there is no degeneracy of the energy in the quantum numbers
$n$, $L$ and $m_{z}$. We now analyze some relationships displayed
in this expression and discuss various limits.

Let us point out the coupling between the rotating frame's angular
velocity $\omega$ with the angular momentum $L$ in the very first
term of Eq. (\ref{eq:E}). This coupling affects the energy eigenvalues
and ultimately the frequency $\nu$ of the related quantum particle.
The effect of rotation on the frequency of light quanta was examined
by Sagnac as early as 1913 \cite{Sagnac1913a,Sagnac1913b}. We noticed
an analogous effect on the bosonic particle due to the rotating frame
in our cosmic string background. This effect was already pointed out
by Ref.~\cite{Castro}. The novelty of our work is to recognize an
additional contribution of the Sagnac-type effect due to non-commutativity.
As a matter of fact, the last term, within the square root of Eq.~(\ref{eq:E}),
couples $\Omega_{3}$ to both $\omega$ and $\left|L\right|$. Moreover,
the cosmic string parameter $\eta$ participates in the coupling.

Eq.~(\ref{eq:E}) contains also a coupling between the angular momentum
and the oscillator's frequency $\omega_{0}$, as well as the non-commutativity
parameter $\Omega_{3}$, as shown by the last term within the square
root. The angular velocity couples with the non-commutativity parameter
through $\left(\Omega_{3}\omega/2\right)^{2}$. Note that there is
no coupling between $\Omega_{3}$ and $\omega_{0}$; the same goes
for $\omega$ and $\omega_{0}$. The non-commutativity parameter $\Omega_{3}$
couples with the string's angle deficit parameter $\eta$ via $\left(\Omega_{3}/2\eta\right)^{2}$
and the last term in Eq. (\ref{eq:E}). Finally, the mass $m$ relates
directly with the oscillator's frequency $\omega_{0}$, through the
combination $m\omega_{0}$, and $L$, but $m$ does not couple with
$\Omega_{3}$ nor $\omega$.

It is helpful to analyze some limits in order to obtain further physical
interpretations. First, we consider the non-rotating limit, $\omega\rightarrow0$,
for which the energy eigenvalues are 
\begin{equation}
\mathcal{E}_{nLm_{z}}=\sqrt{m^{2}-2m\omega_{0}\left(1-m_{z}\right)-\left(\frac{\Omega_{3}}{2\eta}\right)^{2}+2\left(2n+1+\frac{\left|L\right|}{\eta}\right)\sqrt{\left(m\omega_{0}\right)^{2}}}.
\end{equation}
This can be simplified, because $m\omega_{0}>0$, to 
\begin{equation}
\mathcal{E}_{nLm_{z}}=\sqrt{m^{2}+2m\omega_{0}\left(2n+\frac{\left|L\right|}{\eta}+m_{z}\right)-\left(\frac{\Omega_{3}}{2\eta}\right)^{2}},\label{equ68}
\end{equation}
with the following restriction on $\Omega_{3}$: 
\begin{equation}
m^{2}+2m\omega_{0}\left(2n+\frac{\left|L\right|}{\eta}+m_{z}\right)\ge\left(\frac{\Omega_{3}}{2\eta}\right)^{2}.
\end{equation}
Eq. (\ref{equ68}) shows that the non-commutativity term contributes
to the reduction of the oscillator's energy. More specifically, energy
is reduced if the non-commutativity parameter $\Omega_{3}$ is increased
or for smaller values of the angle deficit parameter $\eta$.

Secondly, we discuss the large mass limit. We rewrite Eq. (\ref{eq:E})
as 
\begin{equation}
\mathcal{E}_{nLm_{z}}=-\omega L+m\sqrt{1-2\frac{\omega_{0}}{m}\left(1-m_{z}\right)-\frac{1}{m^{2}}\left(\frac{\Omega_{3}}{2\eta}\right)^{2}+2\left(2n+1+\frac{\left|L\right|}{\eta}\right)\sqrt{\left(\frac{\omega_{0}}{m}\right)^{2}+\left(\frac{\Omega_{3}}{2}\frac{\omega}{m^{2}}\right)^{2}}},
\end{equation}
so that, for a large mass $m$, we consider all terms in the square
root (except the number 1) small compared to 1 ($\sqrt{1-x}\approx1-\frac{x}{2}$),
so that 
\begin{equation}
\mathcal{E}_{nLm_{z}}\simeq-\omega L+m\left[1-\frac{\omega_{0}}{m}\left(1-m_{z}\right)-\frac{1}{2m^{2}}\left(\frac{\Omega_{3}}{2\eta}\right)^{2}+\left(2n+1+\frac{\left|L\right|}{\eta}\right)\sqrt{\left(\frac{\omega_{0}}{m}\right)^{2}+\left(\frac{\Omega_{3}}{2}\frac{\omega}{m^{2}}\right)^{2}}\right],
\end{equation}
and after a few further steps, 
\begin{equation}
\mathcal{E}_{nLm_{z}}\approx m+\omega_{0}\left(2n+\frac{\left|L\right|}{\eta}+m_{z}\right)-\omega L+\frac{\Omega_{3}^{2}}{8m}\left[\left(2n+1+\frac{\left|L\right|}{\eta}\right)\frac{\omega^{2}}{m\omega_{0}}-\left(\frac{1}{\eta}\right)^{2}\right].\label{equ69}
\end{equation}
This shows that the effect of the non-commutative term depends on
the sign of its coefficient, 
\begin{equation}
\left[\left(2n+1+\frac{\left|L\right|}{\eta}\right)\frac{\omega^{2}}{m\omega_{0}}-\left(\frac{1}{\eta}\right)^{2}\right].
\end{equation}
If this term is positive (resp. negative), then a larger non-commutativity
parameter $\Omega_{3}$ will increase (resp. decrease) the energy
level.

For large $m$, if we keep only the dominant terms in Eq. (\ref{equ69})
and neglect the last term in $\frac{\Omega_{3}^{2}}{8m}$, 
\begin{equation}
\mathcal{E}_{nLm_{z}}\approx m+\omega_{0}\left(2n+\frac{\left|L\right|}{\eta}+m_{z}\right)-\omega L,\label{E_non_rel_Castro}
\end{equation}
then we find that the non-commutativity does not contribute. This
result recovers Eq.~(70) in Castros's paper \cite{Castro}, if we
restrict ourselves to $m_{z}=0$. Eq.~(\ref{E_non_rel_Castro}) contains
the mass of the particle plus the standard non-relativistic harmonic
oscillator energy states plus the Page-Werner term \cite{Page,Werner}
accounting for the coupling between the particle's angular momentum
and the angular velocity of the rotating frame.

Lastly, we discuss the commutative limit, $\Omega_{3}\rightarrow0$.
If this case, Eq. (\ref{eq:E}) reduces to 
\begin{equation}
\mathcal{E}_{nLm_{z}}=-\omega L+\sqrt{m^{2}-2m\omega_{0}\left(1-m_{z}\right)+2\left(2n+1+\frac{\left|L\right|}{\eta}\right)\sqrt{\left(m\omega_{0}\right)^{2}}}.
\end{equation}
With $m\omega_{0}>0$, it can be further simplified, 
\begin{equation}
\mathcal{E}_{nLm_{z}}=\sqrt{m^{2}+2m\omega_{0}\left(2n+\frac{\left|L\right|}{\eta}+m_{z}\right)}-\omega L.\label{E_non_commut}
\end{equation}
This equation is similar to Eq. (69) in Ref. \cite{Castro} with the
identification $k_{z}^{2}=2m\omega_{0}m_{z}$.


\subsection{$\mathcal{E}$ for a finite hard-wall \label{SubSec:HDfinite}}

We must emphasize that the energy quantization discussed in Section
\ref{SubSec:HDinf} is meaningful only if the hard-wall $\rho_{0}\equiv1/\omega\eta$
is taken in the limit $\rho_{0}\rightarrow\infty$, otherwise, the
wave function should be zero at the hard-wall. This requirement works
as a boundary condition which is satisfied if the Kummer function
is zero at $\rho_{0}$. Hereafter we analyze this condition and its
physical interpretation in more details. Since it is not possible
to express the zeros of the Kummer function as a closed expression
of its arguments, we need to perform a numerical analysis, which we
compare to section 4.2 of Ref. \cite{Castro}. Among others, we will
point out the effect of the non-commutativity parameter $\Omega_{3}$.

Our numerical analysis involves the following parameters in our differential
equations and in the Kummer function, Eq. (\ref{equ-Kummer}): $\xi$
defined in Eq. (\ref{eq:xi}), $S$ in Eq. (\ref{eq:S}), $T$ in
Eq. (\ref{eq:T}), $q$ in Eq. (\ref{eq:q_plus}), $\upsilon_{0}^{2}$
in Eq. (\ref{eq:xi0}), $a$ in Eq. (\ref{equ-a}), $b$ in Eq. (\ref{equ-b}),
$\chi$ in Eq. (\ref{equZ}), $\rho_{0}$ in Eq. (\ref{hardwall}).
Hereafter, the label ``${hw}$'' stands for ``hard-wall''. In
the commutative limit, $\Omega_{3}=0$, these expressions become $\xi\equiv\left(m\omega_{0}\right)^{\frac{1}{2}}\rho,$
$\chi=m\omega_{0}\rho^{2},$ and $\chi_{hw}=\frac{m\omega_{0}}{\left(\omega\eta\right)^{2}},$
so that $_{1}F_{1}\left(a;b;\chi_{hw}\right)={}_{1}F_{1}\left(a;b;\frac{m\omega_{0}}{\left(\omega\eta\right)^{2}}\right).$
With a set of numerical values given below, we utilized Mathematica
\cite{Mathematica} to find the roots $a=a_{nL}$ of the Kummer function
$_{1}F_{1}(a;b;\chi_{hw})$, where the subscript $n$ stand for its
$n-$th root, while $L$ indicates the dependence of these roots on
this quantum number. In the non-commutative case, $\Omega_{3}\neq0$,
$\chi=\left[\left(m\omega_{0}\right)^{2}+\left(\frac{\Omega_{3}}{2}\omega\right)^{2}\right]^{\frac{1}{2}}\rho^{2}$
and $\chi_{hw}=\frac{\left[\left(m\omega_{0}\right)^{2}+\left(\frac{\Omega_{3}}{2}\omega\right)^{2}\right]^{\frac{1}{2}}}{\left(\omega\eta\right)^{2}}.$
For the sake of comparison, we choose the same numerical factors as
in Ref. \cite{Castro}: $\omega_{0}=0.1$, $\omega=0.5$ and $1.0$,
$\eta=0.1$, 0.5 and 0.9, $L=0$, 1, and $m=1$.

The roots are related to the $a$ parameter in Eq.~(\ref{equ-a}),
\begin{equation}
2a_{nL}=1+q-\upsilon_{0}^{2}/2,
\end{equation}
where $\upsilon_{0}$ depends on the energy. So we can invert the
above equation to find $\mathcal{E}=\mathcal{E}\left(a_{nL}\right)$:
\begin{equation}
\mathcal{E}_{nLm_{z}}=-\omega L\pm\sqrt{m^{2}-2m\omega_{0}\left(1-m_{z}\right)-\left(\frac{\Omega_{3}}{2\eta}\right)^{2}+\left[-4a_{nL}+2+2\sqrt{\left(\frac{L}{\eta}\right)^{2}}\right]\left[\left(m\omega_{0}\right)^{2}+\left(\frac{\Omega_{3}}{2}\omega\right)^{2}\right]^{\frac{1}{2}}}.
\end{equation}
The next step is to compute several values of $\mathcal{E}_{nLm_{z}}$
by choosing multiple values for the parameters $\left\{ m,\omega_{0},\omega,\eta,\Omega_{3}\right\} $
besides the quantum numbers $\left\{ n,L,m_{z}\right\} $.

\begin{figure}[H]

\begin{centering}
\includegraphics[scale=0.40]{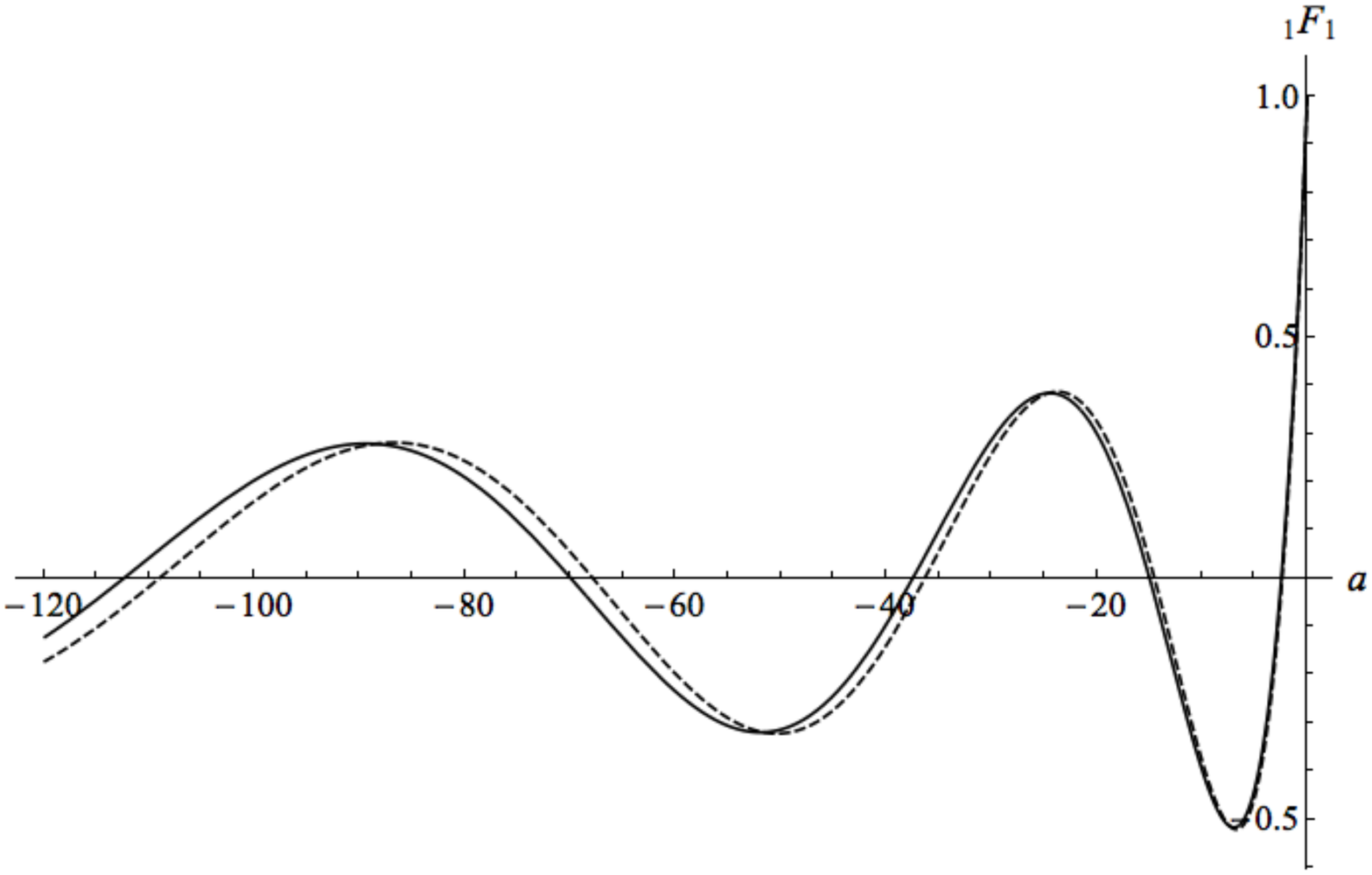} 
\par\end{centering}
\caption{Plot of the zeros $a_{nL}$ ($n=1,2,3$) of the Kummer function $_{1}F_{1}$
for $\Omega_{3}=0$ (solid line) and $\Omega_{3}=0.1$ (dashed line)
for the fixed values of parameters $m=1$, $\omega_{0}=0.1$, $\omega=0.5$,
$\eta=0.9$ and $L=0$.}

\label{Fig-aNL-Omega05Eta09L0} 
\end{figure}

\begin{figure}[H]

\begin{centering}
\includegraphics[scale=0.45]{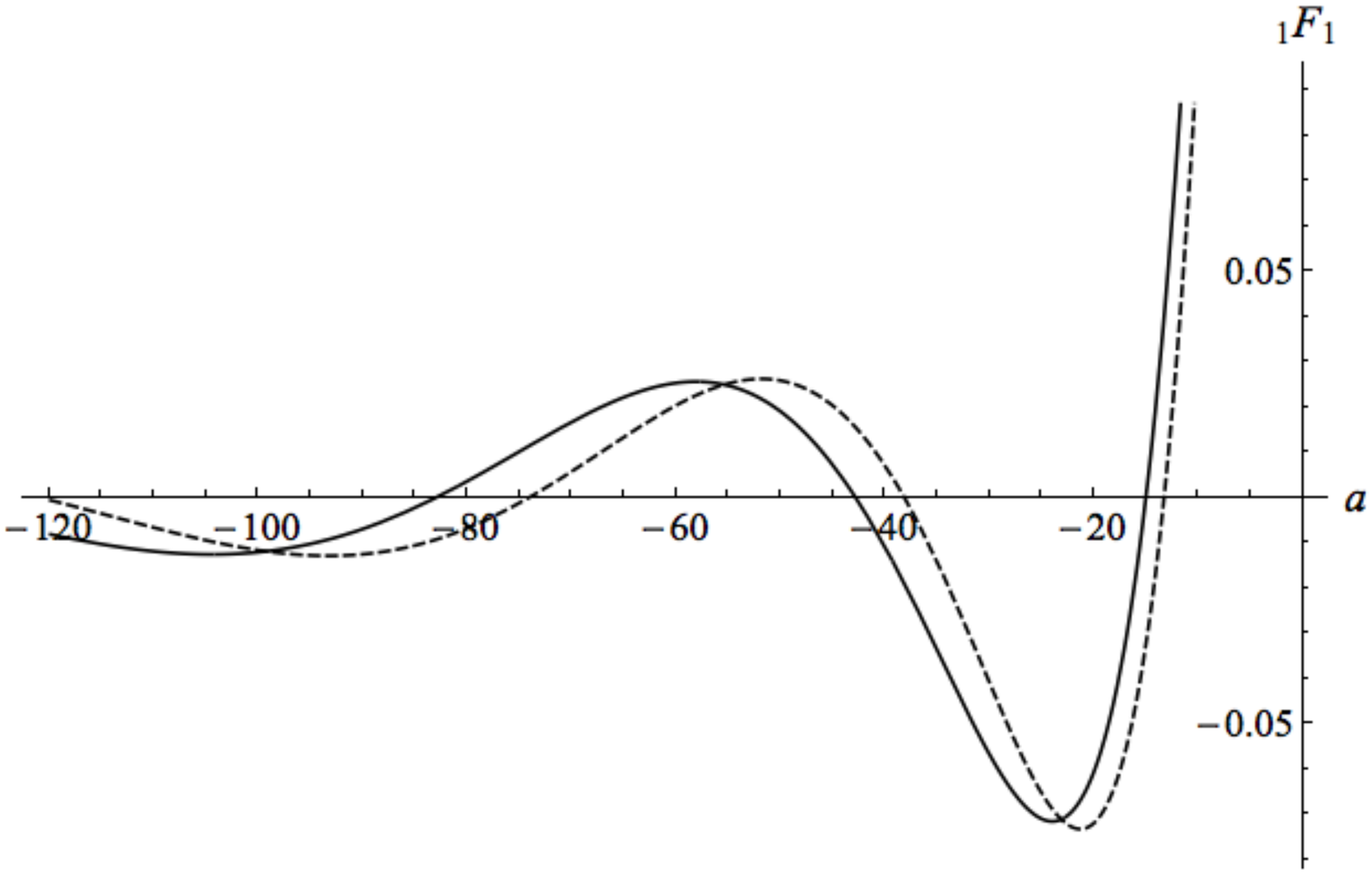} 
\par\end{centering}
\caption{Plot of the zeros $a_{nL}$ ($n=1,2,3$) of the Kummer function $_{1}F_{1}$
for $\Omega_{3}=0$ (solid line) and $\Omega_{3}=0.1$ (dashed line)
for the fixed values of parameters $m=1$, $\omega_{0}=0.1$, $\omega=1.0$,
$\eta=0.5$ and $L=1$.}

\label{Fig-aNL-Omega10Eta05L1} 
\end{figure}

Plots of the roots of the Kummer function are given in Figs.~\ref{Fig-aNL-Omega05Eta09L0}
and \ref{Fig-aNL-Omega10Eta05L1}, which display the zeros $a_{nL}$
for a few chosen values of our parameter space plotted for the commutative
case, $\Omega_{3}=0$, (solid curve) and the non-commutative case,
$\Omega_{3}\neq0$, (dashed curve). The plots in Figs. 1 and 2 show
that the absolute values of the roots $a_{nL}$ of the Kummer polynomials
are reduced by the presence of non-commutativity.

Tables 1 and 2 correspond to the tables 1 and 2 in Ref. \cite{Castro}.
Castro considered $n=1,\cdots,5$ whereas we took $n=1,2,3$, since
we had at least two additional parameters to analyse: $\Omega_{3}$
and $m_{z}$. Our tables elucidate the influence of the cosmic string
parameter $\eta\ \in(0,1]$, related both to the string's linear mass
density and to the deficit angle, on the zeros $a_{nL}$ and on the
energy eigenvalues ${\mathcal{E}}_{\pm}$. 

\begin{table}[H]
\begin{centering}
\begin{tabular}{|c|c|c|c|c|c|c|c|}
\cline{3-8} 
\multicolumn{1}{c}{} &  & \multicolumn{6}{c|}{$L=0$}\tabularnewline
\cline{3-8} 
\multicolumn{1}{c}{} &  & \multicolumn{3}{c|}{$\Omega_{3}=0$} & \multicolumn{3}{c|}{$\Omega_{3}=0.1$}\tabularnewline
\hline 
\multirow{2}{*}{$\eta$} & \multirow{2}{*}{$n$} & \multirow{2}{*}{$a_{nL}$ } & \multicolumn{2}{c|}{$\left|\mathcal{E}_{\pm}\right|$} & \multirow{2}{*}{$a_{nL}$} & \multicolumn{2}{c|}{$\left|\mathcal{E}_{\pm}\right|$}\tabularnewline
\cline{4-5} \cline{7-8} 
 &  &  & $m_{z}=0$  & $m_{z}=1$  &  & $m_{z}=0$  & $m_{z}=1$ \tabularnewline
\hline 
\multirow{3}{*}{0.9} & $1$  & $-2.4546$  & $1.4078$  & $1.4771$  & $-2.3680$  & $1.4069$  & $1.4763$ \tabularnewline
\cline{2-8} 
 & $2$  & $-14.9645$  & $2.6431$  & $2.6806$  & $-14.5051$  & $2.6427$  & $2.6802$ \tabularnewline
\cline{2-8} 
 & $3$  & $-37.4516$  & $3.9976$  & $4.0225$  & $-36.3209$  & $3.9973$  & $4.0223$ \tabularnewline
\hline 
\multirow{3}{*}{0.5} & $1$  & $-0.4896$  & $1.0935$  & $1.1814$  & $-0.4652$  & $1.0899$  & $1.1781$ \tabularnewline
\cline{2-8} 
 & $2$  & $-4.3861$  & $1.6596$  & $1.7188$  & $-4.2478$  & $1.6576$  & $1.7168$ \tabularnewline
\cline{2-8} 
 & $3$  & $-11.3311$  & $2.3521$  & $2.3942$  & $-10.9858$  & $2.3507$  & $2.3928$ \tabularnewline
\hline 
\multirow{3}{*}{0.1} & $1$  & $0.0000$  & $1.0000$  & $1.0954$  & $0.0000$  & $0.8696$  & $0.9778$ \tabularnewline
\cline{2-8} 
 & $2$  & $-1.0000$  & $1.1832$  & $1.2649$  & $-1.0000$  & $1.0810$  & $1.1698$ \tabularnewline
\cline{2-8} 
 & $3$  & $-2.0000$  & $1.3416$  & $1.4142$  & $-2.0000$  & $1.2573$  & $1.3345$ \tabularnewline
\hline 
\end{tabular}
\par\end{centering}
\bigskip{}
\resizebox{\textwidth}{!}{%
\begin{tabular}{|c|c|c|c|c|c|c|c|c|c|c|c|}
\cline{3-12} 
\multicolumn{1}{c}{} &  & \multicolumn{10}{c|}{$L=1$}\tabularnewline
\cline{3-12} 
\multicolumn{1}{c}{} &  & \multicolumn{5}{c|}{$\Omega_{3}=0$} & \multicolumn{5}{c|}{$\Omega_{3}=0.1$}\tabularnewline
\hline 
\multirow{2}{*}{$\eta$} & \multirow{2}{*}{$n$} & \multirow{2}{*}{$a_{nL}$} & \multicolumn{2}{c|}{$\mathcal{E}_{+}$} & \multicolumn{2}{c|}{$\mathcal{E}_{-}$} & \multirow{2}{*}{$a_{nL}$} & \multicolumn{2}{c|}{$\mathcal{E}_{+}$} & \multicolumn{2}{c|}{$\mathcal{E}_{-}$}\tabularnewline
\cline{4-7} \cline{9-12} 
 &  &  & $m_{z}=0$  & $m_{z}=1$  & $m_{z}=0$  & $m_{z}=1$  &  & $m_{z}=0$  & $m_{z}=1$  & $m_{z}=0$  & $m_{z}=1$\tabularnewline
\hline 
\multirow{3}{*}{0.9} & $1$  & $-7.0092$  & $1.5065$  & $1.5557$  & $-2.5065$  & $-2.5557$  & $-6.7710$  & $1.5060$  & $1.5552$  & $-2.5060$  & $-2.5552$\tabularnewline
\cline{2-12} 
 & $2$  & $-25.0484$  & $2.8528$  & $2.8825$  & $-3.8528$  & $-3.8825$  & $-24.2715$  & $2.8525$  & $2.8822$  & $-3.8525$  & $-3.8822$\tabularnewline
\cline{2-12} 
 & $3$  & $-53.0832$  & $4.2387$  & $4.2598$  & $-5.2387$  & $-5.2598$  & $-51.4693$  & $4.2385$  & $4.2596$  & $-5.2385$  & $-5.2596$\tabularnewline
\hline 
\multirow{3}{*}{0.5} & $1$  & $-2.7840$  & $1.0854$  & $1.1473$  & $-2.0854$  & $-2.1473$  & $-2.6659$  & $1.0835$  & $1.1455$  & $-2.0835$  & $-2.1455$\tabularnewline
\cline{2-12} 
 & $2$  & $-9.7150$  & $1.7991$  & $1.8422$  & $-2.7991$  & $-2.8422$  & $-9.3890$  & $1.7977$  & $1.8409$  & $-2.7977$  & $-2.8409$\tabularnewline
\cline{2-12} 
 & $3$  & $-19.7364$  & $2.5487$  & $2.5813$  & $-3.5487$  & $-3.5813$  & $-19.1107$  & $2.5476$  & $2.5803$  & $-3.5476$  & $-3.5803$\tabularnewline
\hline 
\multirow{3}{*}{0.1} & $1$  & $0.0000$  & $1.2321$  & $1.2888$  & $-2.2321$  & $-2.2888$  & $0.0000$  & $1.1786$  & $1.2372$  & $-2.1786$  & $-2.2372$\tabularnewline
\cline{2-12} 
 & $2$  & $-1.0000$  & $1.3439$  & $1.3973$  & $-2.3439$  & $-2.3974$  & $-1.0000$  & $1.2972$  & $1.3520$  & $-2.2972$  & $-2.3520$\tabularnewline
\cline{2-12} 
 & $3$  & $-2.0006$  & $1.4494$  & $1.5001$  & $-2.4494$  & $-2.5001$  & $-2.0003$  & $1.4085$  & $1.4602$  & $-2.4085$  & $-2.4602$\tabularnewline
\hline 
\end{tabular}}\caption{Values for the roots $a_{nL}$ of the Kummer function $M\left(a_{nL},\frac{L}{\eta},Z_{\text{hw}}\right)=0$
with the fixed values $m=1,$ $\omega_{0}=0.1$ and $\omega=0.5$.}
\end{table}

\bigskip{}

\begin{table}[H]
\begin{centering}
\begin{tabular}{|c|c|c|c|c|c|c|c|}
\cline{3-8} 
\multicolumn{1}{c}{} &  & \multicolumn{6}{c|}{$L=0$}\tabularnewline
\cline{3-8} 
\multicolumn{1}{c}{} &  & \multicolumn{3}{c|}{$\Omega_{3}=0$} & \multicolumn{3}{c|}{$\Omega_{3}=0.1$}\tabularnewline
\hline 
\multirow{2}{*}{$\eta$} & \multirow{2}{*}{$n$} & \multirow{2}{*}{$a_{nL}$} & \multicolumn{2}{c|}{$\left|\mathcal{E}_{\pm}\right|$} & \multirow{2}{*}{$a_{nL}$} & \multicolumn{2}{c|}{$\left|\mathcal{E}_{\pm}\right|$}\tabularnewline
\cline{4-5} \cline{7-8} 
 &  &  & $m_{z}=0$  & $m_{z}=1$  &  & $m_{z}=0$  & $m_{z}=1$ \tabularnewline
\hline 
\multirow{3}{*}{0.9} & $1$  & $-11.2177$  & $2.3425$  & $2.3848$  & $-9.9821$  & $2.3419$  & $2.3842$ \tabularnewline
\cline{2-8} 
 & $2$  & $-61.2139$  & $5.0483$  & $5.0681$  & $-54.7008$  & $5.0481$  & $5.0679$ \tabularnewline
\cline{2-8} 
 & $3$  & $-151.1562$  & $7.8398$  & $7.8525$  & $-135.148$  & $7.8397$  & $7.8524$ \tabularnewline
\hline 
\multirow{3}{*}{0.5} & $1$  & $-3.1363$  & $1.5015$  & $1.5667$  & $-2.7572$  & $1.4989$  & $1.5642$ \tabularnewline
\cline{2-8} 
 & $2$  & $-18.5757$  & $2.9035$  & $2.9377$  & $-16.5688$  & $2.9023$  & $2.9365$ \tabularnewline
\cline{2-8} 
 & $3$  & $-46.3368$  & $4.4198$  & $4.4424$  & $-41.3994$  & $4.4190$  & $4.4416$ \tabularnewline
\hline 
\multirow{3}{*}{0.1} & $1$  & $-0.0004$  & $1.0001$  & $1.0955$  & $-0.0001$  & $0.8796$  & $0.9867$ \tabularnewline
\cline{2-8} 
 & $2$  & $-1.0238$  & $1.1872$  & $1.2687$  & $-1.0109$  & $1.1071$  & $1.1940$ \tabularnewline
\cline{2-8} 
 & $3$  & $-2.2262$  & $1.3750$  & $1.4458$  & $-2.1337$  & $1.3145$  & $1.3885$ \tabularnewline
\hline 
\end{tabular}
\par\end{centering}
\bigskip{}
\resizebox{\textwidth}{!}{%
\begin{tabular}{|c|c|c|c|c|c|c|c|c|c|c|c|}
\cline{3-12} 
\multicolumn{1}{c}{} &  & \multicolumn{10}{c|}{$L=1$}\tabularnewline
\cline{3-12} 
\multicolumn{1}{c}{} &  & \multicolumn{5}{c|}{$\Omega_{3}=0$} & \multicolumn{5}{c|}{$\Omega_{3}=0.1$}\tabularnewline
\hline 
\multirow{2}{*}{$\eta$} & \multirow{2}{*}{$n$} & \multirow{2}{*}{$a_{nL}$} & \multicolumn{2}{c|}{$\mathcal{E}_{+}$} & \multicolumn{2}{c|}{$\mathcal{E}_{-}$} & \multirow{2}{*}{$a_{nL}$} & \multicolumn{2}{c|}{$\mathcal{E}_{+}$} & \multicolumn{2}{c|}{$\mathcal{E}_{-}$}\tabularnewline
\cline{4-7} \cline{9-12} 
 &  &  & $m_{z}=0$  & $m_{z}=1$  & $m_{z}=0$  & $m_{z}=1$  &  & $m_{z}=0$  & $m_{z}=1$  & $m_{z}=0$  & $m_{z}=1$\tabularnewline
\hline 
\multirow{3}{*}{0.9} & $1$  & $-31.0448$  & $2.6933$  & $2.7202$  & $-4.6933$  & $-4.7202$  & $-27.6583$  & $2.6930$  & $2.7200$  & $-4.6930$  & $-4.7200$\tabularnewline
\cline{2-12} 
 & $2$  & $-103.2043$  & $5.5195$  & $5.5348$  & $-7.5195$  & $-7.5348$  & $-92.1997$  & $5.5193$  & $5.5347$  & $-7.5193$  & $-7.5347$\tabularnewline
\cline{2-12} 
 & $3$  & $-215.3445$  & $8.3467$  & $8.3573$  & $-10.3467$  & $-10.3574$  & $-192.501$  & $8.3465$  & $8.3572$  & $-10.3465$  & $-10.3572$\tabularnewline
\hline 
\multirow{3}{*}{0.5} & $1$  & $-15.0250$  & $1.7221$  & $1.7586$  & $-3.7221$  & $-3.7586$  & $-13.2896$  & $1.7210$  & $1.7575$  & $-3.7210$  & $-3.7575$\tabularnewline
\cline{2-12} 
 & $2$  & $-42.8174$  & $3.3043$  & $3.3275$  & $-5.3043$  & $-5.3275$  & $-38.1468$  & $3.3035$  & $3.3267$  & $-5.3035$  & $-5.3267$\tabularnewline
\cline{2-12} 
 & $3$  & $-82.9228$  & $4.8795$  & $4.8965$  & $-6.8795$  & $-6.8965$  & $-74.0178$  & $4.8790$  & $4.8960$  & $-6.8790$  & $-6.8960$\tabularnewline
\hline 
\multirow{3}{*}{0.1} & $1$  & $-1.3258$  & $\mathbf{0.8789}$  & \textbf{$\mathbf{0.9314}$}  & $-2.8789$  & $-2.9314$  & $-0.9500$  & $\mathbf{0.8532}$  & $\mathbf{0.9064}$  & $-2.8532$  & $-2.9064$\tabularnewline
\cline{2-12} 
 & $2$  & $-4.3039$  & \textbf{$\mathbf{1.1729}$}  & \textbf{$\mathbf{1.2185}$}  & $-3.1729$  & $-3.2185$  & $-3.55911$  & $1.1451$  & $\mathbf{1.1912}$  & $-3.1451$  & $-3.1912$\tabularnewline
\cline{2-12} 
 & $3$  & $-7.8222$  & $1.4757$  & $1.5157$  & $-3.4757$  & $-3.5157$  & $-6.6770$  & $1.4486$  & $1.4891$  & $-3.4486$  & $-3.4891$\tabularnewline
\hline 
\end{tabular}}\caption{Values for the roots $a_{nL}$ of the Kummer function $M\left(a_{nL},\frac{L}{\eta},Z_{\text{hw}}\right)=0$
with the fixed values $m=1,$ $\omega_{0}=0.1$ and $\omega=1.0$.}
\end{table}

For fixed values of $\left\{ n,L,m_{z},\Omega_{3}\right\} $, the
larger the $\eta$, the smaller the absolute value of both $a_{nL}$
and ${\mathcal{E}}_{\pm}$. If instead, we consider fixed values of
the parameters $\left\{ n,L,m_{z},\Omega_{3},\eta\right\} $, then
the absolute values of $a_{nL}$ and ${\mathcal{E}}_{\pm}$ increase
when the parameter $\omega$ goes from $\omega=0.5$ to $\omega=1.0$.

We observe that the effect of non-commutative parameter $\Omega_{3}$
is to decrease the absolute values of $a_{nL}$ and ${\mathcal{E}}_{\pm}$
(for fixed values of the remaining parameters). If we increase the
value of $m_{z}$, then the absolute values of ${\mathcal{E}}_{\pm}$
increase. The values of $a_{nL}$ do not depend on $m_{z}$ and remain
unchanged. Notice that the hard-wall position $\rho_{0}$ does not
depend on $m_{z}$.

We note that the roots $a_{nL}$ and the energies ${\mathcal{E}}_{\pm}$
increase in absolute value as $n$ increases. Finally, let us mention
that the effect of $L$ is to increase the absolute value of $a_{nL}$
and ${\mathcal{E}}_{\pm}$ for the great majority of the cases reported
in Tables 1 and 2. There are only a few exceptions for $\omega=1.0$;
they are highlighted in bold-face numbers in Table 2. This shows that,
in general, there is no standard global pattern for the effect of
quantum number $L$ on both the roots roots $a_{nL}$ and the energies
${\mathcal{E}}_{\pm}$.

Lastly, we point out that there is consistency between the numerical
case of a finite hard-wall and the analytical case in which the hard-wall
approaches infinity. Recall that in the latter case, $a_{nL}=-n$.
In fact, consider the numerical results for $a_{nL}$ and $\eta=0.1$
in Table 1 for which $\omega=0.5$: it is evident that $a_{nL}$ assumes
negative integer values with four digits precision for $L=0$ regardless
of the non-commutative parameter. For $L=1$, this is also true up
to the third decimal place. These remarks show that $\rho_{0}=1/\omega\eta=20$
(for $\omega=0.5$ and $\eta=0.1$) is already a good approximation
to $\rho_{0}\rightarrow\infty$ up to the decimal figures considered
in Table 1. This pattern is even more pronounced here than in the
commutative case \cite{Castro} due to the effect of $\Omega_{3}$
on $a_{nL}$.\bigskip{}

Figs. 3
and 4 show graphic representations of the information within Tables
1 and 2.

\begin{figure}[H]
\begin{centering}
\includegraphics[scale=0.33]{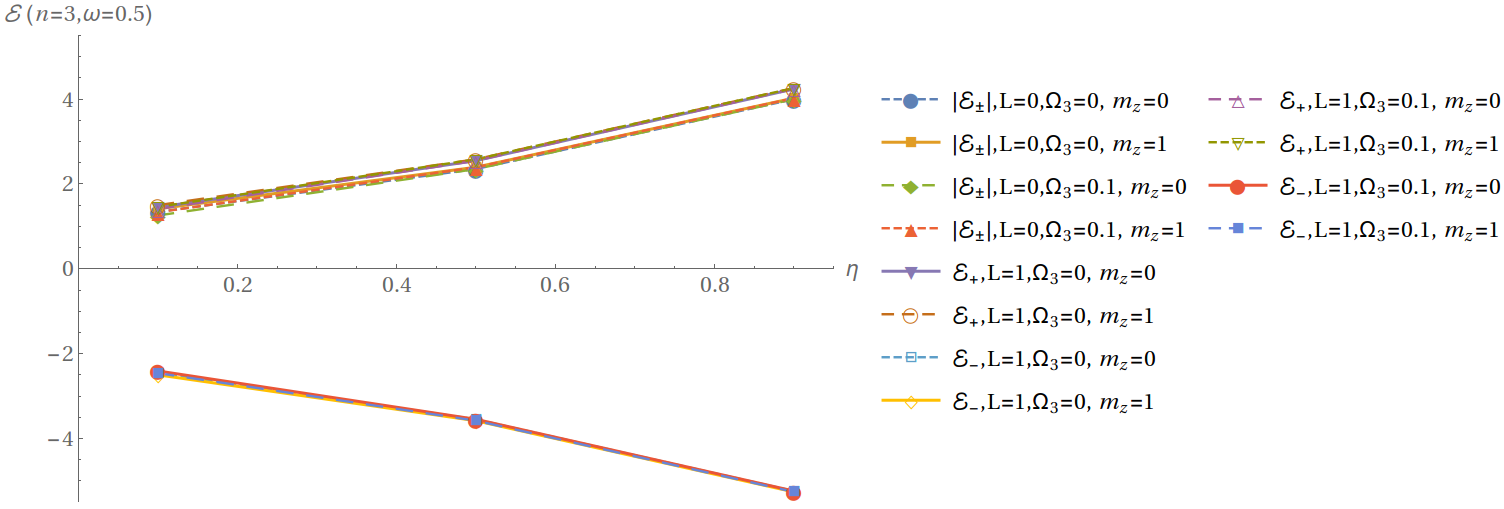}(a)$\quad$
\par\end{centering}
\begin{centering}
\includegraphics[scale=0.33]{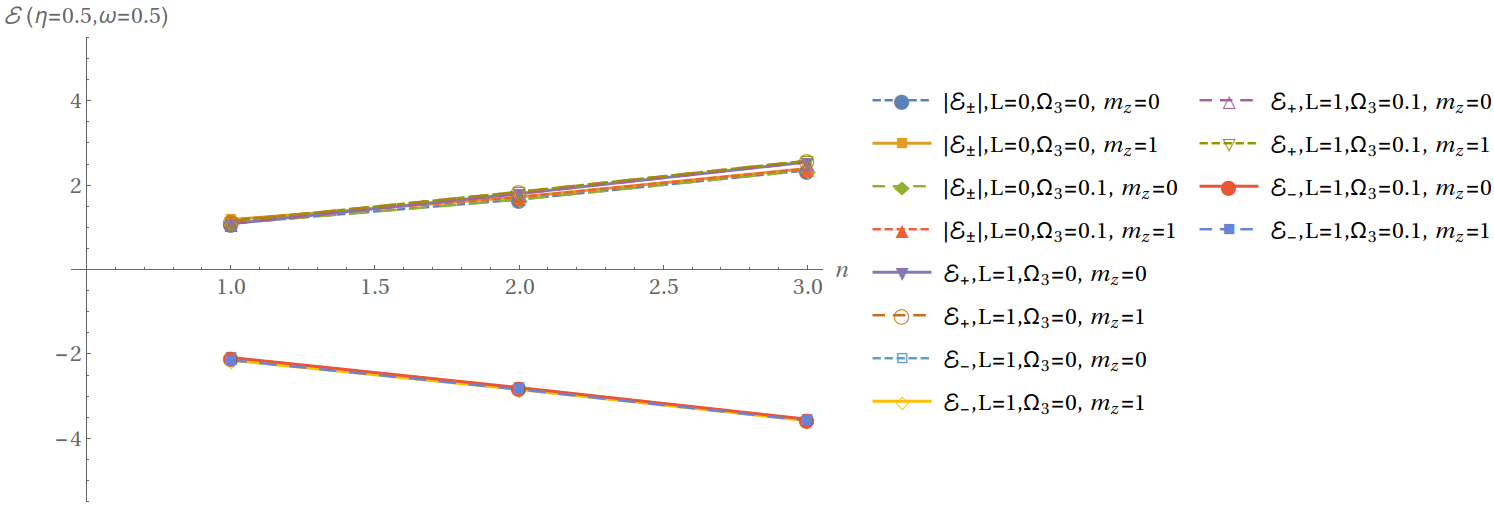}(b)$\quad$
\par\end{centering}
\begin{centering}
\includegraphics[scale=0.35]{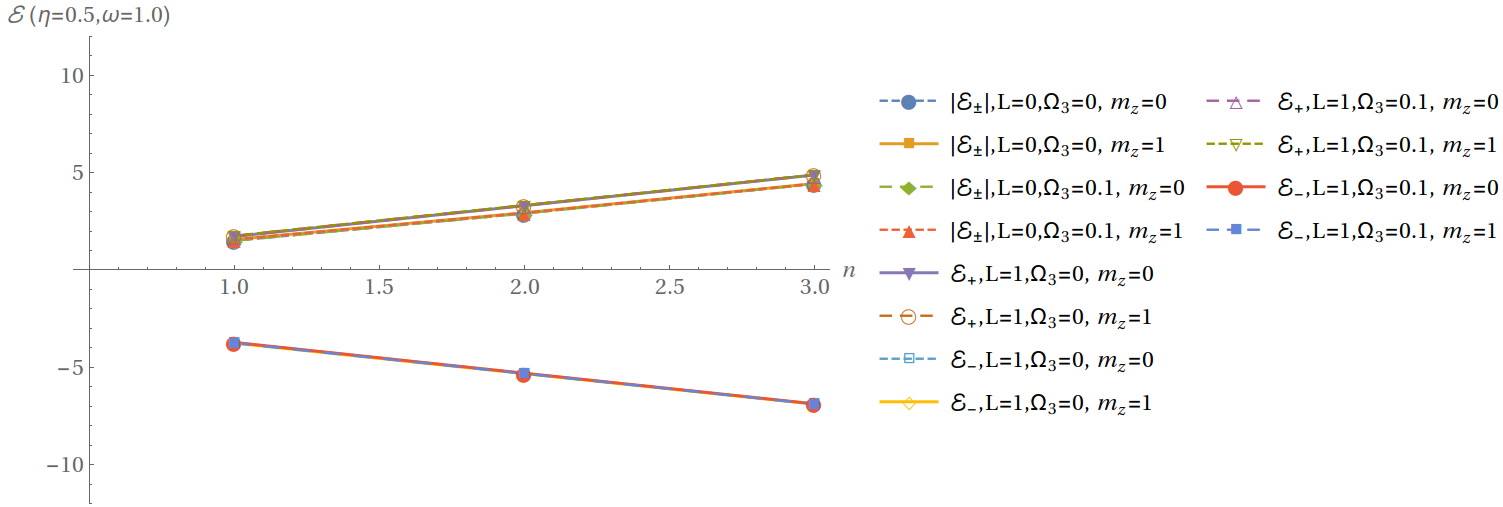}(c)$\quad$
\par\end{centering}
\caption{Energy eigenvalues ${\mathcal{E}}_{\pm}$ (a) as a
function of $\eta$ for $n=3$ and $\omega=0.5$; (b) as a function
of $n$ for $\eta=0.5$ and $\omega=0.5$; and, (c) as a function
of $n$ for $\eta=0.5$ and $\omega=1.0$. }
\label{Fig-E(eta,n,omega)}
\end{figure}

Fig. \ref{Fig-E(eta,n,omega)} displays the effect
of the parameters $\eta$, $n$ and $\omega$ on the eigenvalues $\mathcal{E}_{\pm}$.
In Part (a), we fix $n=3$ and set $\omega=0.5$ while allowing $\eta=0.1,0.5,0.9$.
It enables us to conclude that the absolute value of $\mathcal{E}_{\pm}$
increases with $\eta$. In part (b) we take $\omega=0.5$, $\eta=0.5$
and vary the integer $n$ from 1 to 3. The curves shows that $\left|\mathcal{E}_{\pm}\right|$
increases with the principal quantum number $n$. In part (c), we
plot $\mathcal{E}_{\pm}\left(n\right)$ with $\eta=0.5$ and $\omega=1.0$.
By comparison of parts (b) and (c) of Fig. \ref{Fig-E(eta,n,omega)},
we notice that $\left|\mathcal{E}_{\pm}\right|$ increases for increasing
values of $\omega$. 

\begin{figure}[H]
\begin{centering}
\includegraphics[scale=0.5]{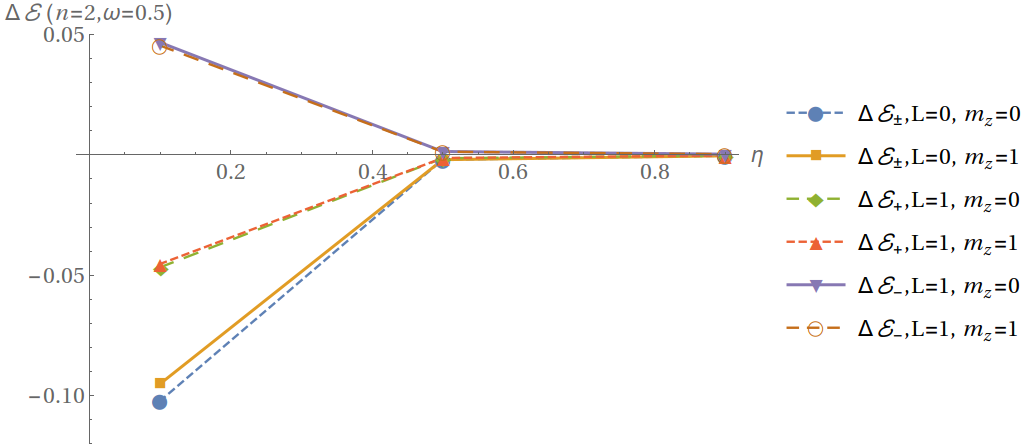}(a)$\quad$
\par\end{centering}
\begin{centering}
\includegraphics[scale=0.5]{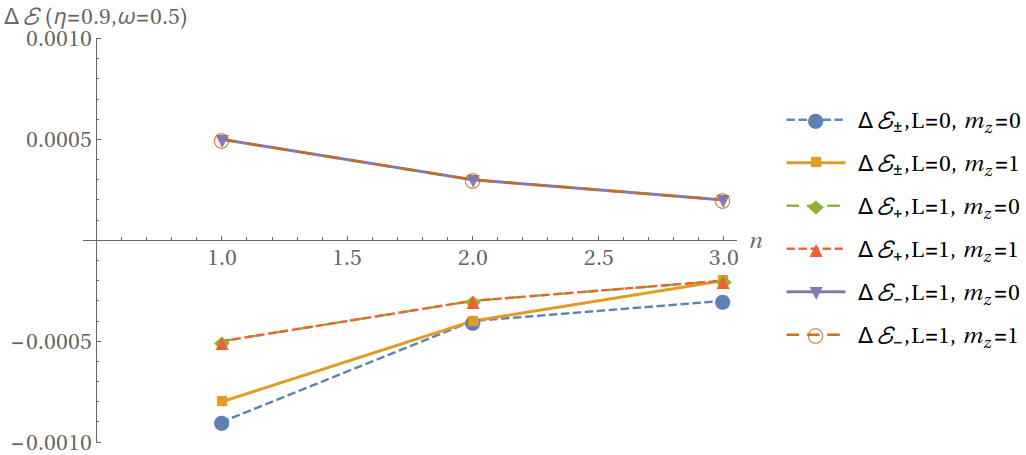}(b)$\quad$
\par\end{centering}
\caption{Plot of the energy difference $\Delta\mathcal{E}_{\pm}\equiv\mathcal{E}_{\pm}\left(\Omega_{3}=0.1\right)-\mathcal{E}_{\pm}\left(\Omega_{3}=0\right)$
in terms of the string parameter $\eta$ and the principal quantum
number $n$: (a) $\Delta\mathcal{E}_{\pm}\left(\eta\right)$ with
$n=2$; (b) $\Delta\mathcal{E}_{\pm}\left(n\right)$ with $\eta=0.9$.
Both graphs are evaluated with $\omega=0.5$.}

\label{Fig-DeltaE(Omega3)}
\end{figure}

Fig. \ref{Fig-DeltaE(Omega3)} presents the effect
of non-commutativity upon our physical system. 

The energy difference $\Delta\mathcal{E}_{\pm}$ is
defined as the difference of energy calculated at $\Omega_{3}=0.1$
and the corresponding energy at $\Omega_{3}=0$ (commutative case)
for fixed values of $n,\,\eta,\,\omega,\,L$ and $m_{z}$, i.e. $\Delta\mathcal{E}_{\pm}\equiv\mathcal{E}_{\pm}\left(\Omega_{3}=0.1\right)-\mathcal{E}_{\pm}\left(\Omega_{3}=0\right)$.
In part (a) of Fig. \ref{Fig-DeltaE(Omega3)} we show how the values
of $\Delta\mathcal{E}_{\pm}$ vary with $\eta$ when $n=2$ and $\omega=0.5$
(the several lines correspond to the trends for different values of
the remaining parameters). We observe that $\Delta\mathcal{E}_{\pm}$
increases inasmuch the values of $\eta$ decrease. In Fig. \ref{Fig-DeltaE(Omega3)}(b),
we display $\Delta\mathcal{E}_{\pm}\left(n\right)$ while setting
$\eta=0.9$ and $\omega=0.5$. In this case, the energy difference
values decrease with rising values of the quantum number $n$.

To conclude, we are in a position to put into perspective
the effect of the coupling prescription which introduces non-commutativity.
Let us compare Fig. \ref{Fig-DeltaE(Omega3)}
of this paper with the corresponding Fig. 4 of Paper 1, Ref. \cite{CQG1}, which
 shows how differently the non-commutative parameter $\Omega_{3}$
impacts the DKP-like oscillator in comparison with the KG oscillator.
In fact, by comparing the $y$-axis values of the plot in Fig. 4(b)
of this paper against the corresponding $y$-axis values of the plot
in Fig. \ref{Fig-DeltaE(Omega3)}(b) of Ref. \cite{CQG1}, we see
that the values of $\left|\Delta\mathcal{E_{\pm}}\left(n\right)\right|$
for the DKP-like oscillator are one order of magnitude smaller than
the values of $\left|\Delta\mathcal{E_{\pm}}\left(n\right)\right|$
for the KG oscillator. Thus we are led to conclude that the
effect of $\Omega_{3}$ is much more pronounced upon the KG oscillator
than it is upon the DKP-like oscillator. In other words, the
coupling prescription  in Eq. (\ref{nonComm_KG}) used for the KG oscillator
is more powerful than the coupling prescription $\mathbf{p}\rightarrow\ \mathbf{p}-\frac{\boldsymbol{\Omega}\times\mathbf{r}}{2}$  to both $\phi$ and $\phi^*$  used for the DKP-like oscillator when it comes down to changing the
energy spectrum via non-commutativity.


\section{Concluding remarks\label{Sec:Conclusion}}

Spin-zero
fields can be described by either a KG field or a DKP field. The KG
field is a solution to a second-order differential equation whereas
 DKP field satisfies a first-order differential equation.
Both KG and DKP descriptions for the spin-zero field are equivalent
in the absence of interactions. This equivalence in not necessarily
established in the presence of interaction (e.g. with the introduction
of the oscillator prescription). In the face of the possibility of
non-equivalence, it is fair to ask the question: What is the more
adequate description of the spin-zero field? The answer calls for
the characterization of both KG and DKP, their different predictions,
phenomenology, and coupling prescriptions. This paper studied of the
coupling prescription related to a DKP-type of oscillator where the
possibility of non-commutativity is also considered. 

One of the questions that guided our investigation
was the following: Is there a second-order differential equation for
the spin-zero field that is (different from the KG description but)
equivalent to the DKP formulation? The answer to this question is
positive as long as we adopt a coupling prescription for the spin-zero
oscillator that is different from the one in Ref. \cite{CQG1}. The
difference is materialized through distinct sign choices in the coupling,
which were be specified in Section \ref{SecCosmicString}. 

The introduction of non-commutativity in the momentum
space of the DKP-like oscillator leads to new phenomenology, in the
sense that a novel wave function is obtained as solution to the field
equation with its very own energy spectrum. This was thoroughly discussed
here. Moreover, we were able to show the equivalence to Castro\textquoteright s
DKP oscillator of Ref. \cite{Castro} in the commutative limit.

Here, we continued work done in Part 1 (in Ref.\cite{CQG1})
and examined the spin-zero DKP-like oscillator in a rotating frame
with a cosmic string space-time endowed with non-commutative geometry
in the momentum space, in order to assess the effects of these geometrical
structures. As far as we know, our two papers are the first investigation
of this generalized Klein-Gordon equation with non-commutative geometry.
We looked for time-independent solutions in cylindrical coordinates
and restricted the study to one non-zero non-commutativity parameter,
$\Omega_{3}$, along the cosmic string axis. We introduced the oscillator
in analogy with the Dirac oscillator of Ref. \cite{Moshinsky}, and
solved the resulting three-dimensional equation by implementing the
separation of variables in terms of radial distance $\rho$, azimuth
$\varphi$ and axial coordinate $z$.

The azimuthal function takes the familiar form in terms of the angular
momentum quantum number $L$ of Eq. \eqref{eq:QuantizationL}, whereas
the axial coordinate sector is, unsurprisingly, expressed in terms
of the Hermite polynomials, which are familiar solutions of the quantum
harmonic oscillator. The most intricate sector of the solution concerns
the radial coordinate which, after considering the asymptotic behaviours
at the origin and at infinity, involves the confluent hypergeometric
function of the first kind, or Kummer function. From these radial
solutions, we obtained the energy quantization by taking into account
the hard-wall condition in the limit where $\rho_{0}=\frac{1}{\omega\eta}$
approached infinity, as well as for a finite value of $\rho_{0}$.
For the hard-wall at infinity, we obtained an expression for the energy
eigenvalues in terms of a principal quantum number $n$ as well of
various physical parameters. We discussed various limits for these
parameters and observed various couplings between physical quantities.
Among others, we noticed that the non-commutativity parameter $\Omega_{3}$
may reduce or enhance the energy values, depending on other parameters.
For the hard-wall at finite distance from the cosmic string, we performed
a numerical search of the roots of the Kummer function in terms of
various parameters. We were able to compare our results, in the commutative
limit, with Ref. \cite{Castro} whose author also studied the non-inertial
effects on the dynamics of scalar boson, but by solving the Duffin-Kemmer-Petiau
oscillator. Among many other observations, we noticed that the non-commutativity
parameter $\Omega_{3}$ reduces the absolute value of the energy.
Since we solved the three-dimensional spatial problem, we thus extend
the analysis in Ref. \cite{Castro}, including in the commutative
limit. Other papers on similar topics restrict their analysis to the
plane $z=0$, including the energy quantization. We did not impose
this restriction nor its additional implications on the wave function
and energy spectrum.

Finally, we compare the results of the present paper with those of
Part 1. This allows us to analyze the role of the coupling prescriptions
in each case (Eqs.\eqref{nonminimal_KG},\eqref{nonComm_KG} in Part
1 and Eqs.\eqref{nonminimal},\eqref{pNC} in Part 2). Several points
can be highlighted. Firstly, we compare the axial part of the solution,
$Z(z)$: In Part 1, a Gaussian function is obtained as solution, while
here, the Gaussian function is multiplied by a Hermite polynomial.
This is a significant difference, since in the latter, the separation
constant is quantized; in the former, the separation constant $K$
remains arbitrary. Concerning the radial solution, the definition
of the dimensionless variable $\xi\equiv S\rho$, with $S$ given
by Eq. \eqref{eq:S} in the present paper, led to an equation that
could be solved analytically, without approximations. The solution
ended up being, essentially, a Gaussian multiplied by a Kummer function.
In Part 1, the same definition for the dimensionless variable resulted
in a dissipative system with an imaginary rest mass in the non-relativisitic
limit. There, this problem was solved with a different choice of dimensionless
variable, where the non-commutative parameter took no part in. The
result was an equation that could be solved only perturbatively, leading
to a solution given by a Gaussian multiplied by a Whittaker function.
As a consequence, the resulting energy presented a very distinct behaviour.
The analytical expression for the quantized energy in Part 1 showed
that the absolute value of energy decreases with the quantum number
$n$; here, this same quantity increases with this parameter. When
the numerical values for the quantized energy were obtained, the effect
of the non-commutative parameter in Part 1 was to increase the absolute
value of energy in comparison with the commutative case; here, the
energy decreases. All these differences stemmed from the coupling
prescription adopted for each case.

Our initial interest in the behaviour of a spin-zero oscillator in
cosmic string backgrounds stems from the influence of topological
defects in cosmology. In the context of particle physics, our investigation
of the Klein-Gordon equation clearly helps to describe the motion
of scalar bosons, such as pions, in the presence of a cosmic string.
Our results can be also of interest in condensed matter physics, given
the general connections between particle physics phenomena (such as
topological defects, cosmic strings or spontaneous symmetry breaking)
and linear topological defects in solids, such as (edge, screw) dislocations
or (wedge, twist) disclinations. Non-commutative geometry has had
a resurgence of interest among particle physicists about twenty years
ago, mainly due to a paper by Seiberg and Witten who argued that the
coordinate functions of the endpoints of strings constrained to a
D-brane in the presence of a constant Neveu-Schwarz B-field would
satisfy a non-commutative algebra \cite{NC,NC2,NC3}. The work \cite{Susskind}
by Bigatti and Susskind considers non-commutativity in the context
of string theory in the presence of a D3-brane and a constant large
magnetic field. Afterwards there were studies of the quantum Hall
effect on non-commutative spaces \cite{Hall,Hall2,Hall4}, so that
possible applications in condensed matter systems include the quantum
Hall effect for interacting bosons as a example of a symmetry-protected
topological phase, as in Refs. \cite{Levin,Sterdyniak,Senthil}, although
considering the effects of topological defects and non-commutative
geometry. Finally, another class of applications could concern Bose-Einstein
condensation along the lines of Refs. \cite{Casana,LMAbreu} which
employed the formalism of the Duffin-Kemmer-Petiau theory (equivalent
to KG for spinless field) in the relativistic and Galilean cases,
respectively. For instance, the angular velocity necessary for a rotating
Bose-Einstein condensate for the formation of vortices, as in Refs.
\cite{Baym,Pethick} could be influenced by the non-commutativity
parameter.




\section*{Acknowledgements}

RRC extends his gratitude to CNPq-Brazil for partial financial support (grant number 309984/2020-3) and to the Instituto Tecnol\'ogico de Aeron\'autica (SP, Brazil) for its hospitality. MdeM is grateful to the Natural Sciences and Engineering Research Council (NSERC) of Canada for partial financial support (grant number RGPIN-2016-04309), to the Instituto Tecnol\'ogico de Aeron\'autica (SP, Brazil) and the Universidade Federal de Alfenas, Campus Po\c cos de Caldas (MG, Brazil) for their hospitality. The authors thank the reviewers for helpful comments that helped them to produce a more robust paper in terms of physical insights.

\end{document}